\documentclass[showkeys,twocolumn,showpacs,preprintnumbers,amsmath,amssymb,prl,10pt,aps]{revtex4-1}
\usepackage{graphicx}
\begin{document}
\preprint{APS/123-QED}
\title{Inhomogeneity and transverse voltage in superconductors}

\author{A. Segal}

\author{M. Karpovski}
\author{A. Gerber}
\affiliation{Raymond and Beverly Sackler School of Physics and Astronomy, Tel Aviv University, Ramat Aviv 69978, Tel Aviv, Israel}

\begin{abstract}
Voltages parallel and transverse to electric current in slightly inhomogeneous superconductors can contain components proportional to the field and temperature derivatives of the longitudinal and Hall resistivities. We show that these anomalous contributions can be the origin of the zero field and even-in-field transverse voltage occasionally observed at the superconductor to normal state transition. The same mechanism can also cause an anomaly in the odd-in-field transverse voltage interfering the Hall effect signal.  
\end{abstract}

\pacs{74.25.F-, 74.25.Wx, 73.50.Jt, 74.78.-w}
\maketitle

\section{Introduction}
   Hall effect in the superconducting state is one of the major tools for studying vortex dynamics, and as such, enjoyed much attention over the last two decades. Two puzzling features observed in the vicinity of the superconductor-normal phase transition attracted particular interest, but nevertheless remain elusive until now: development of an excess transverse voltage at zero applied magnetic field \cite{francavilla_1991,francavilla_1995} and an even-in-field transverse voltage (ETV) under applied field \cite{da_luz_2005,vasek_2004,yamamoto}, and reversal of the Hall coefficient polarity \cite{van_beelen,hagen_1990,nagaoka}.
   
   Appearance of a transverse voltage at zero applied magnetic field was ascribed to attraction among vortices and antivortices, generated at two opposite edges of a film by the self-field of the applied electrical current  \cite{francavilla_1995,glazman}. Attractive vortex-antivortex interaction modifies the vortex trajectory by providing a velocity component parallel or anti-parallel to the current, thus generating a local electric field transverse to current direction. Polarity of this local field depends on the trajectory of individual vortices; therefore, development of the non-zero transverse voltage across macroscopic samples can only take place if the symmetry of the interaction is broken along the current line, a condition which is hard to justify. Breaking of time reversal symmetry due to the fractional statistics in two dimensional high temperature superconductors was suggested \cite{vasek_2004_nofield} as an alternative interpretation, although such a mechanism cannot explain the presence of the effect in conventional three-dimensional superconductors. 
   
   Development of the ETV under applied magnetic field has been mostly discussed in terms of guided vortex motion \cite{pastoriza,danna,marconi,staas_1964,silhanek,wordenweber,basset,soroka_2007}. In this scenario the pinning landscape plays a crucial role. Non-zero ETV can be generated if vortices are forced to move along tracks not-normal to the current direction over a length comparable with the distance between the voltage probes. Such guided motion can be achieved in materials with orientational pinning anisotropy, like single crystals with oriented twin boundaries \cite{pastoriza,danna}, ordered arrays of Josephson junctions \cite{marconi}, foils treated by mechanical rolling \cite{van_beelen,staas_1964}, films with particular lithographic patterning  \cite{silhanek,wordenweber,basset} or deposited onto faceted substrates \cite{soroka_2007}. Surprisingly, both ETV and zero field transverse voltage were also observed in a variety of untreated superconductors without any orientational pinning  \cite{francavilla_1991,francavilla_1995,da_luz_2005,vasek_2004,vasek_2003}, where transport properties should be isotropic. 
   
   In general, the magnitude and details of anomalous transverse voltage effects are unpredictable and hardly reproducible. Samples produced and measured under the same experimental conditions can give a different ETV response \cite{yamamoto}. Similarly inconsistent are reports on the sign reversal of the Hall coefficient. While the effect was reported by many groups in different low and high temperature superconductors  \cite{van_beelen,hagen_1990,nagaoka}, no sign reversal was detected by others in the exact same materials  \cite{galffy}. 
   
   A non-uniform transport current due to inhomogeneity of material was suggested  \cite{francavilla_1991,doornbos} as a possible explanation of at least a part of the anomalous behavior. Against this it was argued  \cite{vasek_2004_nofield} that the effect was observed both in high quality samples with a narrow superconducting transition and in disordered samples with a wide transition. 
   
   An interesting feature noted in several cases  \cite{da_luz_2005,vasek_2004_nofield} is a correlation between the anomalous temperature dependent transverse voltage and the temperature derivative of the longitudinal resistivity $\partial R_{xx}/\partial T$. A similar correlation was also found \cite{yamamoto} between the ETV and the field derivative of the longitudinal resistance $\partial R_{xx}/\partial H$. 
   
   In this article we report on a systematic study of the transverse voltage in several low $T_c$ and high $T_c$ superconducting materials. We shall demonstrate that the anomalous transverse voltage in untreated superconductors can be consistently explained by the presence of a minor asymmetric spatial inhomogeneity in the material, while the correlation with the derivatives of the longitudinal and Hall resistivities is the trademark of its origin.

\begin{figure*}
\includegraphics[width=12cm,height=8cm]{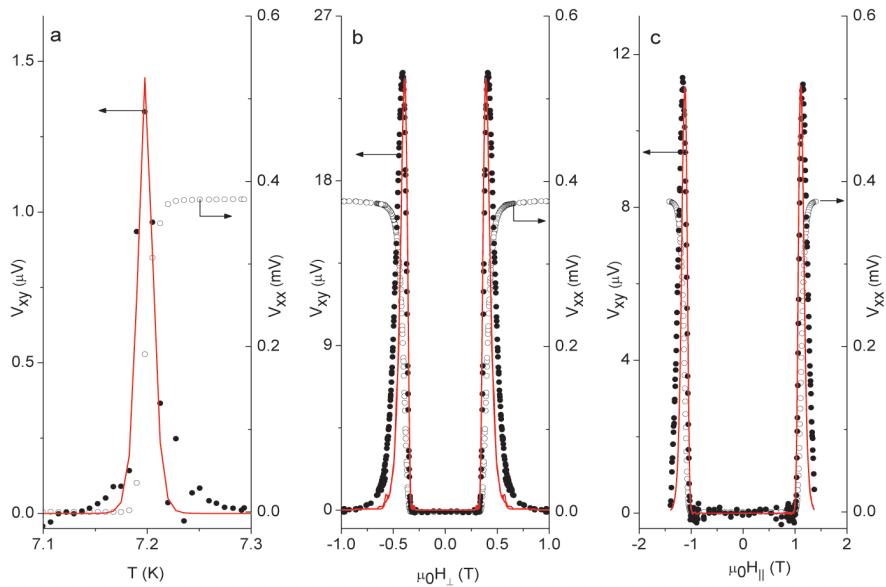}
\caption{\label{fig:one}$V_{xy}$ (\textbullet) and $V_{xx}$ ($\circ$) measured in a 200 nm thick Pb film as a function of  temperature at zero applied magnetic field (a); as a function of field applied normal to the film plane (b) and in-plane parallel to electric current (c) at T = 4.2 K. Solid lines (red online) are fits calculated according to Eqs. \ref{eq:three},\ref{eq:four}.}
\end{figure*}  

\section{Experimental results and discussion}   
   The primary mechanism responsible for depinning and motion of vortices in type II superconductors is the Lorentz force  $\vec{F}_L=\vec{J}\times \vec{B}$, produced by the transport current $\vec{J}$  and magnetic induction $\vec{B}$  per unit volume of the vortex lattice.  The force is largest when magnetic field is applied normal to the electrical current and diminishes to zero when the field is aligned strictly along the current line. As such, any phenomenon related to flux motion is expected to depend strongly on the presence and orientation of applied field relative to electrical current. Fig. \ref{fig:one}a presents the transverse voltage $V_{xy}$, measured as a function of temperature in a plain 200 nm thick Pb film in zero applied magnetic field. The shown $V_{xy}$ signal in this and following figures was obtained by subtracting the normalized mismatch voltage corresponding to an unavoidable misalignment of the transverse contact pads.  The longitudinal voltage measured simultaneously is marked by $V_{xx}$. It can be seen that $V_{xy}$ appears at the onset of the transition and disappears when resistivity reaches its normal state value. Fig. \ref{fig:one}b presents the longitudinal voltage and the ETV measured in the same sample at 4.2 K as a function of field applied normal to the film plane. The data are shown for both magnetic field polarities. Non-zero ETV appears at the onset of the resistive transition and disappears in the normal state. ETV is also observed in the Lorentz force free configuration with field applied in-plane parallel to the current contacts line, shown in Fig. \ref{fig:one}c. Misalignment of the field orientation from the film plane, estimated by the Hall resistance slopes at temperatures above $T_c$, does not exceed 1-2 degrees, which is much too small to justify the observed effect. The magnitude of $V_{xy}$ in both field orientations is very close, implying that the anomalous effect does not depend on the orientation and the very presence of the applied magnetic field. These findings contradict any direct correlation between the transverse voltage and motion of vortices.   

\begin{figure}
\includegraphics[width=8.5cm]{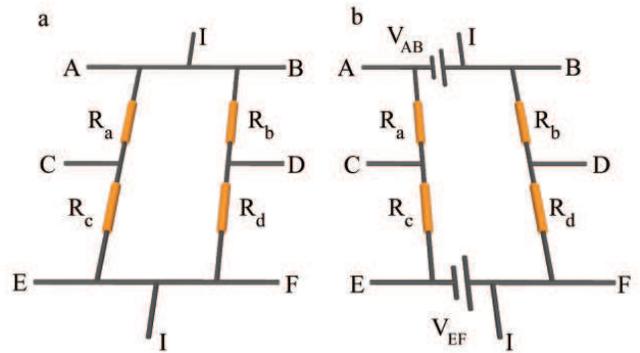}
\caption{\label{fig:two}Model circuits representing a superconducting film: (a) without Hall voltage, (b) with Hall voltage. Each resistor represents a quarter of the sample.}
\end{figure}

   A phenomenon closely related to the subject of this work is an occasional observation of an excess voltage in longitudinal resistivity across the superconductor-normal state transition  \cite{francavilla_1991,santhanam}. Vaglio et al  \cite{vaglio} suggested that the effect can be explained by inhomogeneity of the material and described by a simple current distribution model. Here we adapt and extend this model to treat the development of the transverse voltage. The sample is represented by a four resistors network, shown in Fig. \ref{fig:two}a, where resistors $R_{a-d}$ represent four quarters of a superconducting sample and thus depend on field and temperature. Longitudinal voltage $V_{xx}$ is measured between contacts A and E or B and F, and the transverse voltage $V_{xy}$ is measured between C and D. We are interested in $V_{xx}$ and $V_{xy}$ in the transition range where conductivity of any macroscopic section of the sample is finite. For simplicity we assume all resistors to be Ohmic. $V_{xx}$ and $V_{xy}$ are calculated by use of Kirchhoff's laws as:

\begin{equation}
	\label{eq:one}
	V_{xx}=I\frac{(R_a+R_c)(R_b+R_d)}{R_a+R_b+R_c+R_d}
\end{equation}

\begin{equation}
	\label{eq:two}
	V_{xy}=I\frac{R_aR_d-R_bR_c}{R_a+R_b+R_c+R_d}
\end{equation}

where $I$ is the total electric current. A non-zero transverse voltage $V_{xy}$ will be generated in any case of diagonal inequality $R_aR_d \neq R_bR_c$. We assume the simplest case in which three resistors are identical $R_a(T,H) = R_b(T,H) = R_c(T,H) \equiv R^*(T,H)$ and the fourth $R_d$ passes from the superconducting to the normal state with a small delay in temperature $\Delta T_t$ and/or field $\Delta H_t$, that we will denote as the transverse delays. In the transition range $R_d$ is given by $R_d(T,H)=R^*(T+\Delta T_t,H+\Delta H_t)\approx R^*(T,H)+\Delta T_t\frac{\partial R^*(T,H)}{\partial T}+\Delta H_t\frac{\partial R^*(T,H)}{\partial H}$ . When the superconducting transition is crossed as a function of temperature in zero or constant magnetic field, the transverse voltage $V_{xy}$ will develop according to:
\begin{equation}
	\label{eq:three}
	V_{xy}=\frac{\Delta T_t}{4}\frac{\partial V_{xx}(T,H)}{\partial T}
\end{equation}

Similarly, when magnetic field is varied at constant temperature, the model predicts the transverse voltage to be given by: 

\begin{equation}
	\label{eq:four}
	V_{xy}=\frac{\Delta H_t}{4}\frac{\partial V_{xx}(T,H)}{\partial H}
\end{equation}

$\Delta H_t$ changes sign at the negative field polarity, therefore the resulting $V_{xy}(H)$ is an even function of the magnetic field. $V_{xy}(T,H)$ is proportional to the temperature or/and field derivative of the longitudinal resistance and becomes significant at the superconducting transition due to a sharp variation of resistance. Solid lines in Figs. \ref{fig:one}a-c are fits to Eqs. \ref{eq:three} and \ref{eq:four} calculated using the measured longitudinal resistance and one fitting parameter $\Delta T_t$ or $\Delta H_t$ only. A perfect fit in Fig. \ref{fig:one}a was obtained with $\Delta T_t = 2.8\cdot 10^{-4} K$, which is more than two orders of magnitude smaller than the width of the superconducting transition $\delta T = 5\cdot 10^{-2} K$, the latter being defined as the temperature span over which resistivity changes between $10\%$ and $90\%$ of its normal value. Fits of $V_{xy}(H)$ drawn in Figs. \ref{fig:one}b and \ref{fig:one}c were calculated with $\Delta H_t= 230G$ and $\Delta H_t= 170G$ for the perpendicular and parallel field orientations respectively. The transition widths in these orientations are by an order of magnitude larger: $\delta H_{\bot}  = 2000G$ and $\delta H_{||}= 3000G$. 

\begin{figure}
\includegraphics[width=8.5cm]{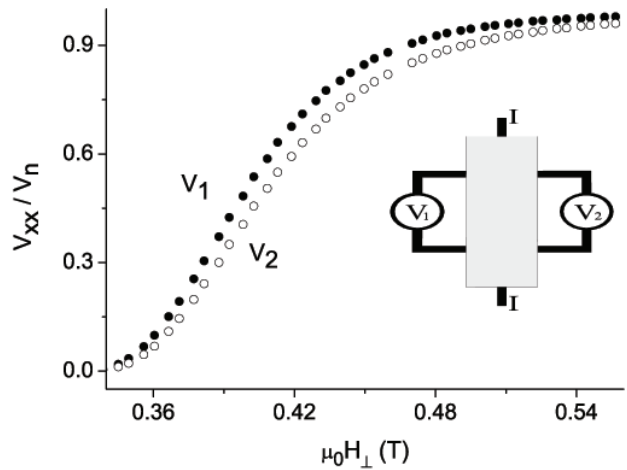}
\caption{\label{fig:three}The normalized longitudinal voltage $V_{xx}/V_n$ measured at two opposite edges of the Pb film as a function of magnetic field applied normal to the film. $V_n$ are the normal state values at 1 T. Inset shows a sketch of the sample and location of the voltage probes.}
\end{figure}

   Presence of the spatial inhomogeneity can be tested explicitly by e.g. simultaneous probing of different parts of a sample. Fig. \ref{fig:three} presents two magnetoresistance measurements taken simultaneously along two opposite edges of Pb film with field applied normal to the film plane. Both sets of data are normalized by their respective normal state values at the field of 1 T, denoted by $V_n$. Inset of Fig. \ref{fig:three} shows a sketch of the sample and location of the voltage probes. $V_1$ in Fig. \ref{fig:three} is identical to $V_{xx}$ in Fig. \ref{fig:one}b. The relative shift of the transition is about $250G$ in a fair agreement with the value $\Delta H_t= 230G$ found by the fitting.  

\begin{figure}
\includegraphics[width=8.5cm]{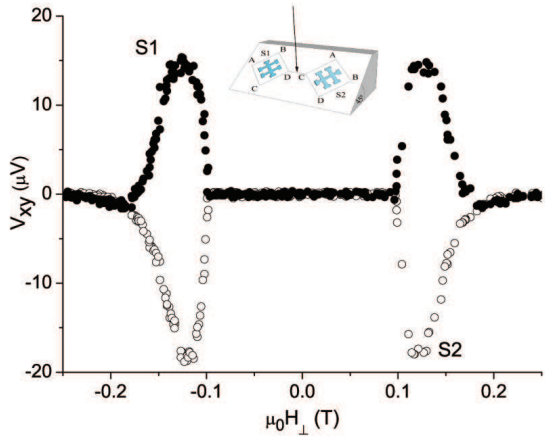}
\caption{\label{fig:four}Transverse voltage measurements of two Pb samples with thickness gradients oriented at $+45^0$ (\textbullet) and $-45^0$ ($\circ$)  to the current line. A sketch of the sample deposition method is shown in the inset. Arrow indicates the direction to the target.}
\end{figure}

   Following the model, the sign of the transverse voltage depends on the sign of $\Delta T_t$ and $\Delta H_t$, i.e. whether the region "d" has higher or lower critical field/temperature than the rest of the sample. To test the relevance of the model we fabricated two samples with artificial "diagonal" non-uniformity.  Two Pb samples were deposited through two identical Hall bar masks onto glass substrates tilted by $45^o$ from the target direction. Masks were rotated in plane of the substrates by $+45^o$ and $-45^o$, as shown in the sketch in Fig. \ref{fig:four}. By using the labels of Fig. \ref{fig:four}, corner B of sample S1 was closer to the target than corner C, and corner A of sample S2 was closer than corner D. As a result, thickness of opposite corners (B, C in S1, and A, D in S2) varied by about $25\%$ for an average thickness of 200 nm, and the thickness gradient was oriented by about $+45^o$ and $-45^o$ from the current line direction. Both samples were prepared simultaneously under the same conditions. For a sample with a variable thickness one expects a lower critical field in the thinnest region due to a higher current density. Fig. \ref{fig:four} presents the transverse voltage $V_{xy}$ measured in the two samples at T=4.2K as a function of field applied normal to the films. The transverse voltage peaks are almost identical but have opposite polarity consistent with the expected: positive for sample S1 and negative for S2. The result was consistently reproduced in a number of similarly fabricated pairs. 

\begin{figure}
\includegraphics[width=8.5cm]{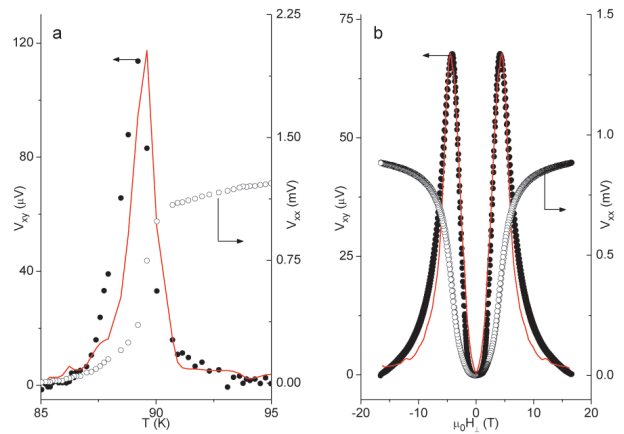}
\caption{\label{fig:five}Even $V_{xy}$ (\textbullet) and $V_{xx}$ ($\circ$) measured in a YBaCuO film as a function of:  (a) temperature at zero applied field; (b) field applied normal to the film plane at T = 81.8 K. Solid lines (red online) are fits calculated according to Eqs. \ref{eq:three},\ref{eq:four}.}
\end{figure}

   Applicability of the model to high temperature superconductors was tested with YBaCuO films grown by off axis dc magnetron sputtering onto Yttrium-stabilized $ZrO_2$ covered sapphire substrates \cite{almog}. Figs. \ref{fig:five}a and \ref{fig:five}b present respectively the temperature dependence of $V_{xy}$ in zero field and the even $V_{xy}$ measured as a function of field applied normal to ab planes at T=81.8K. The anomalous transverse voltages in YBaCuO have the same characteristic features as found in Pb. The solid lines in Figs. \ref{fig:five}a and \ref{fig:five}b were calculated by Eqs. \ref{eq:three} and \ref{eq:four} using the measured $R(T)$ and $R(H)$ and fitting parameters $\Delta T_t = 0.6 K$  and $\Delta H_t = 1.17 T$. Both fitting parameters correspond to about $20\%$ of the respective transition widths. The same model seems therefore to describe consistently the anomalous transverse voltage both in Pb and in YBaCuO. 

\begin{figure}
\includegraphics[width=8.5cm]{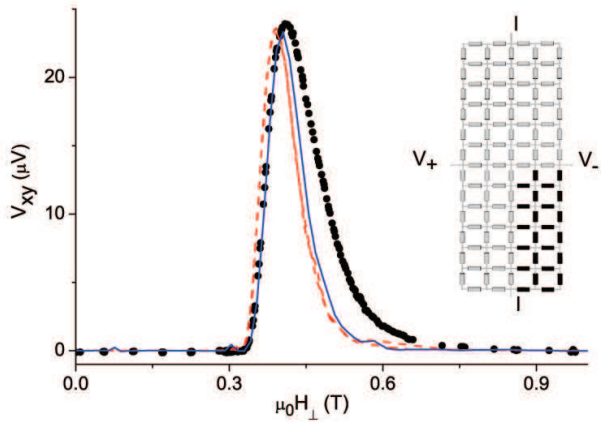}
\caption{\label{fig:six}Even $V_{xy}$ in a Pb sample as function of field applied normal to the film. Full circles represent experimental data, solid line (blue online) is the numerically calculated transverse voltage, and dashed line (red online) is the voltage calculated according to Eq. \ref{eq:four}. Inset shows a sketch of the resistor grid used for the numerical calculation.}
\end{figure}

   The question which needs to be addressed is whether the simple effective four resistors circuit of Fig. \ref{fig:two}a models reliably the current and electric potential distribution in a real sample. In order to test this, we performed a numerical calculation of current flow in inhomogeneous samples. The discrete form of the continuum transport equations is equivalent to current flow in a resistor grid; therefore a natural extension of the four resistors model is its replacement by a large resistor network. We conducted the calculation on a grid consisting of $100\times 300$ resistors depicted in the inset of Fig. \ref{fig:six} (the inset illustrates a grid of $4\times 12$ resistors). Magnetic field dependence of the black resistors is shifted by $\Delta H_t$ relative to the gray resistors. In order to compare the calculation to the experimental results, the field dependence of the gray resistors was taken to be proportional to the measured $V_{xx}(H)$ shown in Fig. \ref{fig:one}b: $R(H)=R_0\cdot V_{xx}(H)/V_{xx}(1T)$ where $V_{xx}(1T)$ is the normal state value of $V_{xx}$ measured at a field of 1 Tesla. Coefficient $R_0$ was determined by demanding the calculated normal state $V_{xx}$ to be equal to the experimental value. Result of the numerical calculation is shown in Fig. \ref{fig:six} by the solid line, with $\Delta H_t=270G$. The numerically calculated curve is close to the one obtained by use of Eq. \ref{eq:four} (dashed line in Fig. \ref{fig:six}) with similar values of $\Delta H_t$ ($270G$ for the numerical calculation compared to $230G$ for Eq. \ref{eq:four}). One can therefore conclude that the simple four resistor model  captures the essential details of the current and electric potential distribution of a macroscopic sample.

\begin{figure}
\includegraphics[width=8.5cm]{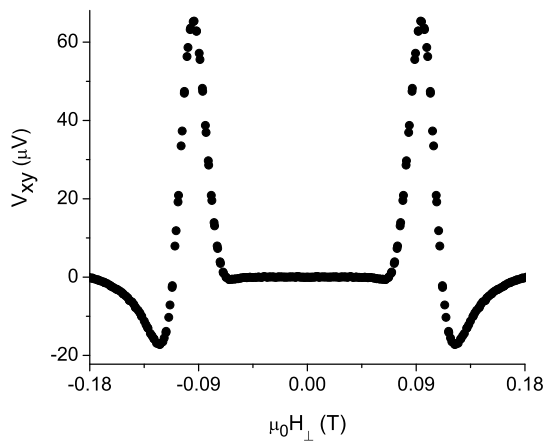}
\caption{\label{fig:seven}Even transverse voltage measured in a 200 nm thick NiPb sample as a function of field normal to the film plane at T=4.2K.}
\end{figure}

   Evidently the inhomogeneity of arbitrary samples can be more complex than modeled above and generate a variety of transverse voltage patterns. One example is the variable polarity ETV found in a granular Ni-Pb mixture with $15\%$ volume of Ni shown in Fig. \ref{fig:seven}. Such a pattern can be explained if e.g. the transition is wider in one quarter of the sample than in others while the critical field, defined at the mid-transition, is roughly the same everywhere. In this case the resistance of the selected region exceeds the rest at one stage of the transition and delays at the other. The ETV, which is determined by the difference between the diagonal resistances, will therefore change sign during the transition. 
   
   The same inhomogeneity mechanism can also be responsible for the generation of an anomalous odd-in-field transverse signal. If the superconducting transition is not identical along the sample, the Hall voltage will also differ at different cross-sections. The representing circuit is shown in Fig. \ref{fig:two}b. $V_{AB}$ and $V_{EF}$ indicate the Hall voltages at two cross-sections, while resistors $R_{a-d}$ model the sample's longitudinal resistance. Assuming that $V_{AB}(H) = V_{EF}(H+\Delta H_l)$ and $R_d=R^*(H+\Delta H_t)$ where $\Delta H_l$ is a small longitudinal delay in the critical field, one can calculate $V_{xx}$ between points F and B, and $V_{xy}$ between points C and D as:

\begin{equation}
	\label{eq:five}
	V_{xx}=IR+\frac{\Delta H_l}{2}\frac{\partial V_{xy}(T,H)}{\partial H}+\frac{\Delta H_t}{4}\frac{\partial V_{xx}(T,H)}{\partial H}
\end{equation}

\begin{equation}
	\label{eq:six}
	V_{xy}=V_{EF}+\frac{\Delta H_l}{2}\frac{\partial V_{xy}(T,H)}{\partial H}+\frac{\Delta H_t}{4}\frac{\partial V_{xx}(T,H)}{\partial H}
\end{equation}

Here in addition to the regular signals $IR$ and $V_{EF}$, both the longitudinal and transverse voltages contain two additional terms: an odd-in-field term proportional to the field derivative of the Hall voltage and an even term proportional to the field derivative of resistivity. These extra "inhomogeneity" terms can be significant when resistivity or Hall voltage vary sharply. Polarity of the odd term $\frac{\Delta H_l}{2}\frac{\partial V_{xy}}{\partial H}$ can be the same or opposite to $V_{EF}$ however its magnitude does not exceed the latter. Therefore, the superposition of two odd terms can result in an anomaly of the Hall effect signal although not a reversal of its polarity. In our samples we did not identify an anomalous odd signal exceeding the experimental accuracy.

   Effect of a minor asymmetric inhomogeneity discussed here is relevant not only for the superconducting transitions but also for other systems where the longitudinal or Hall resistivity vary sharply as a function of any external parameter like pressure, temperature, magnetic or electric field. Specific examples to mention are reversal of magnetization in ferromagnetic films with perpendicular magnetic anisotropy \cite{segal} and development of fractional quantum Hall steps in a 2D electron gas  \cite{pan}.  As a general warning, a straight-forward determination of the longitudinal and Hall resistivities as the even and odd in magnetic field components of the measured data can lead to erroneous conclusions. 

\section{Conclusions}

   In summary, we tested and found no evidence of a direct correlation between the vortex dynamics and development of an anomalous zero field and even in field transverse voltage in untreated low $T_c$ and high $T_c$ superconductors. On the other hand, the effect can be consistently explained by the presence of a minor asymmetric inhomogeneity of the material.  In this case, a simple circuit model predicts an appearance of additional voltage signals proportional to the temperature and field derivatives of the longitudinal resistivity in an excellent agreement with the experiment. The same mechanism can also cause an anomaly in the odd in field transverse voltage interfering the Hall effect signal, although not a reversal of its polarity. The effect can be present both in high quality samples with a narrow and sharp transition, and in disordered samples with a wide transition since only a minor relative transverse inhomogeneity is sufficient for its development.

\section{Acknowledgements}

   The authors acknowledge M. Azoulay, B. Almog and N. Bachar for supplying the YBaCuO samples. This work was supported by the Israel Science Foundation grant No. 633/06.

%

\end{document}